\documentstyle[psfig]{aipproc}

\begin{document}
\def\swb{Schwarzschild~}
\def\lsim{\lower.5ex\hbox{$\; \buildrel < \over \sim \;$}}
\def\gsim{\lower.5ex\hbox{$\; \buildrel > \over \sim \;$}}
\baselineskip 10pt

\title{Frequency Dependent Lags - A Common Phenomenon of Accreting
Sources}
 
\author{X.-M. Hua$^{*,\dagger}$, D. Kazanas$^*$, and W. Cui$^{\ddagger}$}
\address{$^*$LHEA Code 661, GSFC/NASA, Greenbelt, MD 20771\\
$^{\dagger}$Universities Space Research Association, Seabrook, MD 21218\\
$^{\ddagger}$Center for Space Research, MIT, Cambridge, MA 02139}

\maketitle
\vskip -0.5truecm
\begin{abstract}
The Fourier frequency dependent hard X-ray lag, first discovered from
the analysis of aperiodic variability of the light curves of the black
hole candidate Cygnus X-1, turns out to be a property shared by several 
other accreting compact sources. We show that the lag can be explained 
in terms of Comptonization process in coronae of hot electrons with
inhomogeneous density distributions. The density profile of a corona, 
like the optical depth and electron temperature, can
significantly affect the Comptonization energy spectrum from it. This
means, by fitting the energy spectrum alone, it is not possible to
uniquely determine the optical depth and temperature of the
Comptonization cloud if its density profile is unknown. The hard X-ray
time lag is sensitive to the density distribution of the scattering 
corona. Thus simultaneous analysis of the spectral and temporal X-ray 
data will allow us to probe the density structure of the Comptonizing 
corona and thereby the dynamics of mass accretion onto the compact object. 
\end{abstract}
\vskip -0.5truecm
\section*{Introduction}
It is well known that the energy spectrum does not provide sufficient 
information to determine the dynamics of mass accretion in X-ray
binaries. This information can be obtained from the 
independent variability measurements. One of the 
frequently used variability measures is the phase or time lag
of hard X-rays with respect to the soft ones. Because in Comptonization, 
the energy of the escaping photons increases with their residence time 
in the scattering medium, the hard photons lag with respect to softer 
ones by amounts which depend on the photon scattering time. Thus, 
observations of these time lags provide a measure of the local electron 
density, a quantity inaccessible to the spectral analysis alone. 
\begin{figure}[ht]
\centerline{\psfig{file=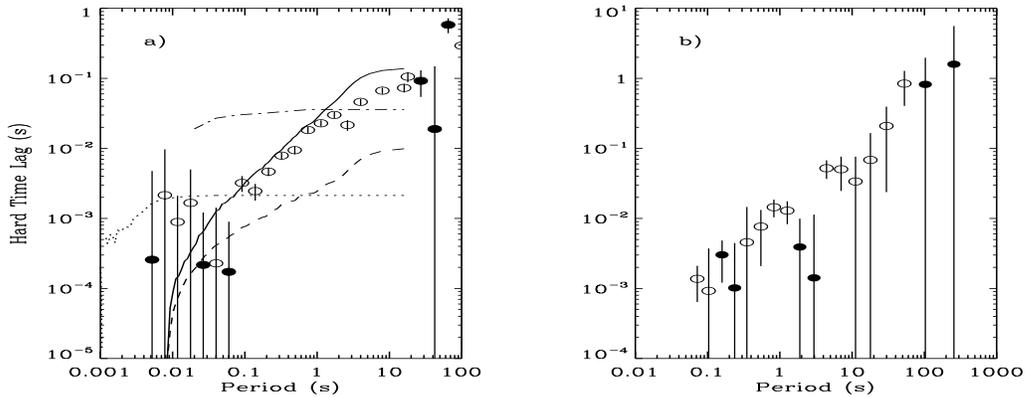,width=5.6in,height=2.2in}}
\caption{\baselineskip 5pt
(a) The time lags of hard X-rays ($15.8 - 24.4$ keV) with respect to 
soft ones ($1.2 -5.7$ keV) obtained from Cyg X-1 data [1]
and (b) that of hard X-rays ($6 - 28$ keV) with 
respect to soft ones ($2 -6$ keV) for the source GRS 1758-258 [4].
The circles and dots represent the measured positive 
and negative lags respectively. The dash-dotted curve represents the lag 
predicted based on analytical Comptonization model 
and assumed corona electron density $10^{16} {\rm cm}^{-3}$.  The three 
other curves are calculation results for Comptonization in the model 
coronae that produce the energy spectra shown in Figure 3a.
}
\end{figure}

Since the high energy radiation is believed to be produced in the 
vicinity of the compact object of the size of a few \swb 
radii, these lags should be roughly of the order $\sim$ msec or shorter. 
Observations of these lags, however, have given results drastically 
different from our expectations: 
The lags measured from the black hole candidate Cyg X-1 obtained by 
Ginga \cite{miya88} has revealed that these lags increase linearly with 
increasing Fourier period to $\Delta t \simeq 0.1$ sec for frequency 
$\omega \simeq 0.1$ Hz (Figure 1a). Similar results were obtained by the 
analysis of the higher quality RXTE data from the high state of the same 
source \cite{cui97b} (Figure 3b). Analysis of the Ginga data from the 
source GX 339-4\cite{miya91} and the RXTE data from 
GRS 1758-258(\cite{smith97}, Figure 1b) gave the similar dependence, 
with $\Delta t$ extending in the latter source to $\sim 1$ sec for 
$\omega \sim 0.02$ Hz! Finally, the high energy transient GRO J0422+32 
was observed by OSSE during outburst and the lags 
indicated a similar behavior (\cite{grove97}, Figure 2a). These results 
suggest that the {\it frequency-dependent hard X-ray time lags may be a 
common property of these sources}.
\begin{figure}[ht]
\centerline{\psfig{file=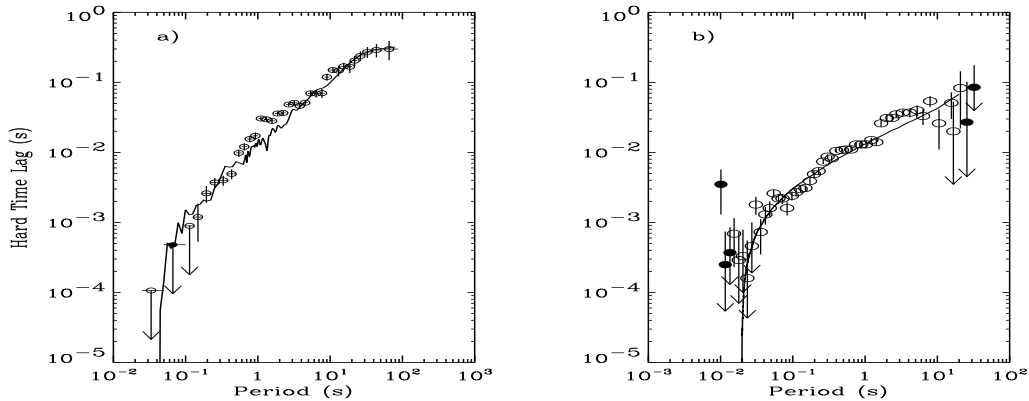,width=5.6in,height=2.2in}}
\caption{\baselineskip 5pt
(a) The time lags of hard X-rays ($75 - 175$ keV) with respect to soft
ones ($35 -60$ keV) for the source GRO~J0422+32, obtained from OSSE 
observation [5]. The solid curve is obtained from a calculation based on 
Comptonization in a model corona with $p=1, r_2 = 6.33$ light seconds, 
$kT_e =100$ keV and $\tau_0 = 0.25$. The level-off frequency 
at $\omega \simeq 0.025$ Hz indicates the edge of the corona 
$r_2 \simeq 1/2\pi \omega$. 
(b) The time lags of hard X-rays ($14.09 - 100$ keV) with
respect to soft ones ($0 -3.86$ keV) for the source Cyg X-1 based on 
RXTE data [9]. The solid curve is a result from the calculation of 
Comptonization in a corona with $p = 3/2, kT_e = 100$ keV, 
$\tau_0 = 1.5$ and $r_2 = 5$ light seconds.}
\end{figure}
\vskip -0.5truecm
\section*{The Extended Atmosphere Model}
Kazanas, Hua \& Titarchuk \cite{kht97} proposed that the discrepancy 
between expectation and observation can be resolved if the the corona of
Comptonizing electrons is non-uniform and extending over
several orders of magnitude in radius. Specifically, they proposed
the following density profile for the Comptonizing medium:
$$n(r) = \cases  {n_1 &for $r \le r_1$ \cr n_1 (r_1/r)^{p} &
for $r_2 > r > r_1$ \cr} \eqno(1)$$
\noindent
where the power index $p$ is a free parameter; $r$ is the radial
distance from the center of the spherical corona; $r_1$ and $r_2$ are
its radii of the inner and outer edges respectively. The density
profile of Eq. (1) allows scattering to take place over a wide range
of densities, thereby introducing time lags over a similar range of
time, leading to the observed $\omega-$dependent lags.

In particular, we
considered density profiles with $p =1$ and 3/2. A soft photon from 
a source located near $r \simeq 0$ would undergo the scattering 
process, which increases its energy and 
introduces hard photon time lags proportional to 
the scattering mean free time in the medium \cite{ht96}. 
The majority of the data available to-date (Figures 1a, 1b, 2a and 3b)
are consistent with the $p =1$ density profile.
The density profile with $p = 3/2$, implied by the 
advection-dominated accretion model suggested by Narayan \& Yi
(\cite{ny94}) would result in a weaker dependence on Fourier period.
There are indications of such a dependence in the 
recent Cyg X-1 data in its hard state (\cite{wilms97}, 
Figure 2b) and during the transitions \cite{cui97b}.
{\it The above arguments, thus, lead us to the notion that 
that these lags can be used to probe the density structure of the
corona.}
\section*{Density Profile and Energy Spectra}
These arguments have been verified by numerical simulation of the
spectral-temporal properties of the Comptonization process
in media with density profile given by Eq. (1) (\cite{kht97}, 
\cite{hkt97}, \cite{hua97}, \cite{hkc97}). The 
simulations also show that the density distribution of the 
Comptonizing electrons can significantly affect the emergent 
spectrum. Figure 3a displays the energy spectra of Comptonizing 
coronae of various density profiles with properly chosen optical 
depths. These coronae give rise to spectra that can fit the 
BATSE data of Cyg X-1 \cite{ling97} equally well. This fact shows 
that fitting the spectra alone does not suffice to determine the 
optical depth $\tau_0$ if the density profile is unknown. 
\begin{figure}[ht]
\centerline{\psfig{file=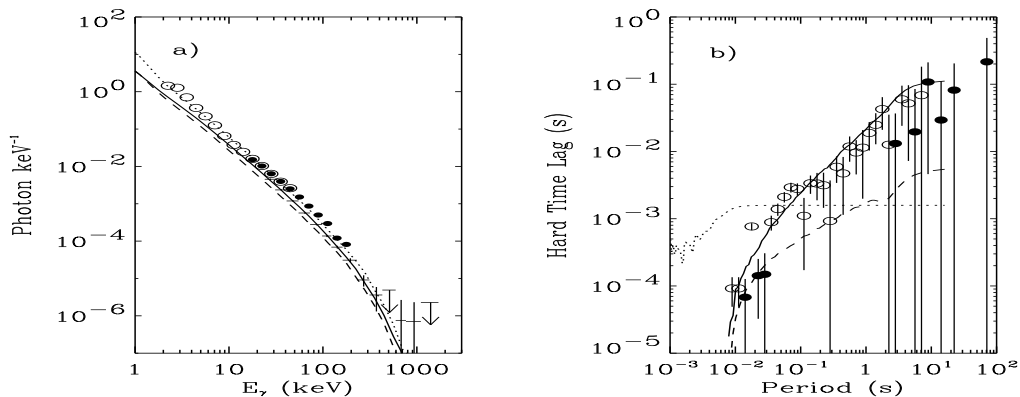,width=5.6in,height=2.2in}}
\caption{\baselineskip 5pt
(a) Three calculated energy spectra which fit equally well the
Cyg X-1 data (crosses) observed by CGRO/BATSE
in 1994 [13]. These spectra result from Comptonization in
coronae with the same temperature (100 keV) but different optical depths
and density profiles: dotted - $p=0, \tau_0=0.5$, solid - $p=1,
\tau_0=1.0$ and dashed - $p=3/2, \tau_0=0.7$. The dotted and dashed
curves are slightly displaced to separate the otherwise nearly
identical curves. Also plotted are RXTE/PCA (circles) and HEXTE (dots)
data from the same source observed in 1996 [14]. Both observations were made
when Cyg X-1 was in the high state and the energy spectra are
consistent with each other except normalization. 
(b) The time lags of hard X-rays ($13 - 60$ keV) with
respect to soft ones ($2 -6.5$ keV) resulting from Comptonization in 
the same coronae that produce the energy spectra shown in Figure 3a. 
The three curves represent the density profiles $p=0, 1$ and 3/2 
respectively. The time lag between the same energy bands based on 
RXTE data from Cyg X-1 [2] are also plotted.}
\end{figure}
On the other hand, the analysis of the lags between
photons in the high and low energy ranges, resulting
from Comptonization in the above three coronae, yields widely different
results (Figure 3b). Also plotted in the figure are the
lags obtained from the RXTE observation \cite{cui97b} whose
corresponding photon spectrum is given in Figure 3a. It is obvious that 
the observation favors the 
non-uniform electron density distributions, specifically the profile 
with $p=1$.

All the fits of the lags in Figures 1a, 2a, 2b and 3b
require that {\it the outer edge of the coronae be at a distance as
great as $r_2 = 5$ light second, or $10^{11}$ cm}! This conclusion 
is rather inescapable, since the observed lags and energy spectra 
require scattering free path of the order 1 light second and $\tau_0 
\sim 1$ respectively. 

It is clear that 1) appropriate modeling of the time properties
of accreting compact sources suggests a very different picture than 
that assumed to-date in modeling their spectra, with direct consequences 
for the associated dynamics and 2) realistic models must 
consider the associated time variability properties of 
the system. The physically meaningful way to model these type
of sources is through combined spectral-temporal models.

\end{document}